\documentclass[twocolumn,showpacs,preprintnumbers,amsmath]{revtex4}

\usepackage[usenames]{color}
\usepackage{graphicx}
\usepackage{epsfig}
\usepackage{dcolumn}
\usepackage{bm}
\usepackage{latexsym}
\usepackage{amsfonts}
\usepackage{amssymb}

\newcommand{\beq}{\begin{equation}}
\newcommand{\eeq}{\end{equation}}
\newcommand{\beqa}{\begin{eqnarray}}
\newcommand{\eeqa}{\end{eqnarray}}
\newcommand{\bea}{\begin{array}}
\newcommand{\ena}{\end{array}}
\begin{document}
\title{The universal area spectrum in single-horizon black holes}
\author{Takashi Tamaki}
\email{tamaki@tap.scphys.kyoto-u.ac.jp}
\affiliation{Department of Physics, Kyoto University, 
606-8501, Japan}
\author{Hidefumi Nomura}
\email{nomura@gravity.phys.waseda.ac.jp}
\affiliation{Department of Physics, Waseda University, 
Ohkubo, Shinjuku, Tokyo 169-8555, Japan}

\date{\today}
\begin{abstract}
We investigate highly damped quasinormal mode of single-horizon 
black holes motivated by its relation to the loop quantum gravity. 
Using the WKB approximation, we show that the real part of the frequency 
approaches the value $T_{\rm H}\ln 3$ for dilatonic black hole 
as conjectured by Medved et al. and Padmanabhan. It is surprising since 
the area specrtum of the black hole determined by the Bohr's 
correspondence principle completely agrees with that of 
Schwarzschild black hole for any values of 
the electromagnetic charge or the dilaton coupling. 
We discuss its generality for single-horizon black holes and the meaning 
in the loop quantum gravity. 
\end{abstract}

\preprint{KUNS-1921}

\maketitle

%%%%%%%%%%%%%%%%%%%%%%%%%
{\it Introduction.} 
%%%%%%%%%%%%%%%%%%%%%%%%%
Progress in the loop quantum gravity (LQG) has been remarkable 
particularly after the introduction of the spin network formalism \cite{Smolin}. 
Due to this formalism, general expressions for the spectrum of the area and the 
volume operators can be derived \cite{Rovelli,Ash1}. For example, 
the area spectrum $A$ is 
%%%%%%%%%%%%%%%%
\beqa
A=8\pi \gamma \sum \sqrt{j(j+1)}\ , \label{AreaLQG}
\eeqa
%%%%%%%%%%%%%%%
where $\gamma$ is the Immirzi parameter related to an ambiguity 
in the choice of canonically conjugate variables \cite{Immirzi}. 
The sum is added up all intersections between a surface and a spin 
network carrying a label $j=0$, $1/2$, $1$, $3/2$, $\ldots$ reflecting 
the SU(2) nature of the gauge group. 
The statistical origin of the black hole entropy $S$ is also derived using 
this formalism (and the introduction of the isolated horizon \cite{Corichi} 
and the U(1) Chern-Simons theory). The result is summarized as 
\cite{Ashtekar} 
%%%%%%%%%%%%%%%%
\beqa
S= \frac{A\ln (2j_{\rm min}+1)}{8\pi \gamma\sqrt{j_{\rm min}
(j_{\rm min}+1)}}\ ,  \label{SLQG}
\eeqa
%%%%%%%%%%%%%%%
where $A$ and $j_{\rm min}$ are the horizon area and the lowest nontrivial representation 
usually taken to be $1/2$ because of SU(2), respectively. In this case, the Immirzi parameter 
is determined as $\gamma =\ln 2/(\pi \sqrt{3})$ to produce the 
Bekenstein-Hawking entropy formula $S=A/4$. This is one of the important 
attainment in the LQG. However, it should be emphasized that progress in 
the LQG is not restricted to theoretical interest. 
Phenomenological role in the early universe and the 
role as a possible source of the Lorentz invariance violation has also 
been discussed \cite{Bojowald}. 

Recently, quite a new encounter to the LQG and the quasinormal mode was 
considered in Ref. \cite{Dreyer}. We explain the idea briefly.  
If we apply the first law of black hole thermodynamics, 
%%%%%%%%%%%%%%%%
\beqa
dA=\frac{4}{T_{\rm H}}dM\ , \label{1st}
\eeqa
%%%%%%%%%%%%%%%
where we only considered the ``infinitesimal" change in gravitational 
mass for simplicity. 
Then we seek for a possibility that there is a lower bound in the area change. 
The discrete area spectrum is also favorable from the observation that 
the horizon area of nonextremal black holes bahaves as a classical 
adiabatic invariant \cite{Bek}, since the Ehrenfest principle says 
that any classical invariant corresponds to a quantum entity 
with discrete spectrum. 
We identify minimum change $dM$ as the real part of the highly damped 
quasinormal mode Re$(\omega )$ 
based on the Bohr's correspondence principle ``transition frequency at large 
numbers should equal classical oscillation frequencies" followed by 
\cite{Hod}. For Schwarzschild black hole, we have \cite{Motl,Andersson} 
%%%%%%%%%%%%%%%%
\beqa
{\rm Re}(\omega )=T_{\rm H}\ln 3 \ {\rm for}\ {\rm Im}(\omega )\to\infty\ . 
\label{Sch}
\eeqa
%%%%%%%%%%%%%%%
In this case, we obtain 
%%%%%%%%%%%%%%%%
\beqa
dA=4\ln 3\ . \label{Spectrum}
\eeqa
%%%%%%%%%%%%%%%
At this point, there is no direct relation to the LQG. 
Interesting and debatable issue is that we identify (\ref{Spectrum}) 
with the minimum area change in the area spectrum (\ref{AreaLQG}), i.e., 
%%%%%%%%%%%%%%%%
\beqa
dA=4\ln 3=8\pi \gamma\sqrt{j_{\rm min}(j_{\rm min}+1)}\ . 
\label{identify}
\eeqa
%%%%%%%%%%%%%%%
By substituting this formula to (\ref{SLQG}), we obtain $j_{\rm min}=1$ 
to produce $S=A/4$. In this case, the Immirzi parameter is modified 
as $\gamma =\ln 3/(2\pi \sqrt{2})$. This consideration calls various 
arguments such as modification of the gauge group SU(2) to SO(3) or 
the modification of the area spectrum in LQG and so on which we will discuss 
later \cite{Alekseev,Corichi2,Kaul,Corichi3,Ling}. 

We must also suspect that only Schwarzschild black hole has the relation 
(\ref{Spectrum}) and the identification (\ref{identify}) has no universality. 
We should notice that the formulae (\ref{AreaLQG}) and (\ref{SLQG}) in the LQG 
do not depend on matter fields since 
their symplectic structures do not have a contribution for the 
horizon surface term \cite{Ashtekar}. Thus, it is important to investigate 
these properties in other black holes in determining whether or not the discussion 
above is related to the LQG. 

The work we should mention are Ref.~\cite{Visser,Pad} which show that the 
imaginary part of the highly damped quasinormal mode have a period proportional 
to the Hawking temperature for the single-horizon black holes. This result suggests 
a generalization of the case in Schwarzschild black hole, i.e.,  
%%%%%%%%%%%%%%%%
\beqa
\omega =T_{\rm H}\ln 3-2\pi T_{\rm H}i\left(n+\frac{1}{2}\right) .  \label{Key}
\eeqa
%%%%%%%%%%%%%%%
For Schwarzschild black hole, this formula applies to scalar and gravitational
perturbations. For electromagnetic perturbations, the real part disappears in this limit. 
What this means in the context of Hod's proposal is not clear at present. 
Their work and Ref~\cite{Cardoso} also suggest that if we are between 
two horizons, we will see a mixed 
contribution from the two horizons. Thus, we cannot see a periodic behavior in 
the imaginary part in general which was also confirmed numerically in 
Ref.~\cite{Shijun} for Schwarzschild-de Sitter black hole. The analysis for 
Reissner-Nordstr\"om black hole in Ref.~\cite{Motl,Andersson} also shows that 
existence of the inner horizon disturbs the imaginary part to be periodic. 
This result agrees with numerical results in Ref.~\cite{Berti}. 
This would also be true for Kerr black hole where the contribution of the 
angular momentum also makes things more complicated \cite{Shijun2}. 

Therefore, the strategy we take here is whether or not the formula (\ref{Key}) holds 
for the single-horizon black holes. From this view point, 
we examine the WKB analysis following Ref.~\cite{Andersson} by exemplifying the case 
for dilatonic black hole \cite{GM-GHS}. (For quasinormal mode of dilatonic 
black hole, see Refs.~\cite{Ferrari}.) Surprisingly, the answer is in the affirmative. 
If one see its derivation, one would confirm the generality for the single-horizon 
black holes. Notice that dilatonic black hole is a charged black hole with 
single-horizon. Thus, considering this model provides the evidence that the 
essential thing that determines whether or not (\ref{Key}) holds is {\it not} 
the electromagnetic charge {\it but} the space-time structure. 
We also consider this direction and thier meaning in the LQG. 

%%%%%%%%%%%%%%%%%%%%%%%%%
{\it The WKB Analysis for single-horizon black holes.}
%%%%%%%%%%%%%%%%%%%%%%%%%
As a background, we consider the static and spherically symmetric metric as 
%%%%%%%%%%%%%%%%
\beqa
ds^{2}=-f(r)e^{-2\delta (r)}dt^{2}+
f(r)^{-1}dr^{2}+r^{2}d\Omega^{2}, 
\label{metric}
\eeqa
%%%%%%%%%%%%%%%%
where $f(r):=1-2m(r)/r$. We define 
%%%%%%%%%%%%%%%%
\beqa
g(r)=e^{-\delta}f(r)\ . \label{benri}
\eeqa
%%%%%%%%%%%%%%%
Notice that \cite{Visser,Pad}
%%%%%%%%%%%%%%%%
\beqa
g'(r_{\rm H})=4\pi T_{\rm H}\ , \label{temperature}
\eeqa
%%%%%%%%%%%%%%%
where $':=d/dr$ and $r_{\rm H}$ is the event 
horizon. Our basic equation for black hole perturbations are 
%%%%%%%%%%%%%%%%
\beqa
\frac{d^2 \psi}{dr_{\ast}^{2}}+[\omega^2 -V(r)]\psi=0\ ,\label{Schro}
\eeqa
%%%%%%%%%%%%%%%
where the time dependence of the perturbations are assumed to be $e^{-i \omega t}$. 
The tortoise coordinate $r_{\ast}$ is defined as 
%%%%%%%%%%%%%%%%
\beqa
\frac{dr_{\ast}}{dr}=\frac{1}{g(r)}\ .
\eeqa
%%%%%%%%%%%%%%%
The potential $V(r)$ for the general case (\ref{metric}) is written 
followed by \cite{Visser,Kar} as 
%%%%%%%%%%%%%%%%
\beqa
V(r)&=&g\left[\frac{l(l+1)}{r^2}e^{-\delta}+(1-k^2)
\frac{2m}{r^3}e^{-\delta}+\right. \nonumber \\
&& \left.(1-k)(\frac{g'}{r}-\frac{2m}{r^3}e^{-\delta})\right]\ .
\eeqa
%%%%%%%%%%%%%%%
For $k=0$, $1$ and $2$, $V(r)$ corresponds to the case for 
the scalar, electromagnetic and the odd parity 
gravitational perturbations, respectively. At present, we cannot obtain 
the form like (\ref{Schro}) for 
the even parity mode. First, we concentrate on the odd parity gravitational 
perturbations, i.e., $k=2$.  
We also define 
%%%%%%%%%%%%%%%%
\beqa
\Psi=g^{1/2}\psi\ .
\eeqa
%%%%%%%%%%%%%%%
Using (\ref{benri}), our basic equation can be rewritten as 
%%%%%%%%%%%%%%%%
\beqa
\Psi''+R(r)\Psi=0\ , \label{basic}
\eeqa
%%%%%%%%%%%%%%%
where 
%%%%%%%%%%%%%%%%
\beqa
R(r)=g^{-2}[\omega^2 -V+(g')^{2}/4-gg''/2]\ .\label{r}
\eeqa
%%%%%%%%%%%%%%%

Then, we consider the WKB analysis combined with the complex-integration technique 
which is a good approximation in the limit Im($\omega )\to -\infty$. 

First, we summarize the analysis for Schwarzschild black hole 
and consider in the complex r-plane below. 
Two WKB solutions in (\ref{basic}) can be written as 
%%%%%%%%%%%%%%%%
\beqa
\Psi_{1,2}^{(s)} (r)=Q^{-1/2}\exp \left[\pm i\int_{s}^{r} Q(x)dx\right]\ , \label{WKB}
\eeqa
%%%%%%%%%%%%%%%
where $Q^{2}=R+$extra term. Here, the extra term is chosen for $\Psi$ to bahave 
near the origin appropriately. From (\ref{basic}), $\Psi (r)\sim r^{1/2\pm 2}$ at 
$r\to 0$. Since $R\sim -15/4r^2$ at $r\to 0$ in Schwarzschild black hole, 
we should choose $Q^{2}:=R-1/(4r^{2})$ for the WKB solution (\ref{WKB}) 
to behave correctly. 

%%%%%%%%%%%%%%%%%%%%
\begin{figure}[htbp]
\psfig{file=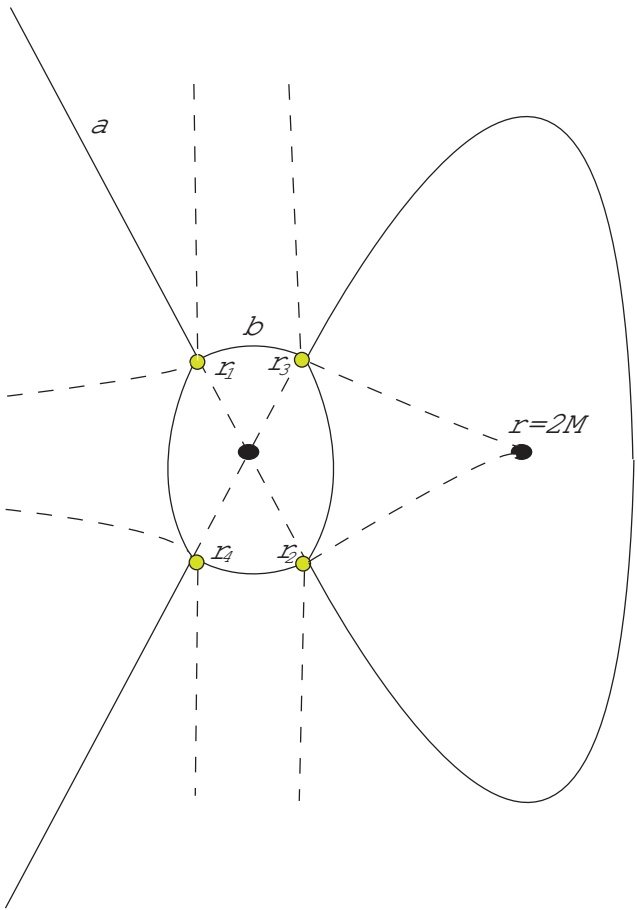,width=3.3in}
\caption{Zeros and poles of $Q^{2}(r)$ and for Schwarzschild black hole in the 
complex $r$-plane in the limit Im$(\omega )\to -\infty$. 
The related Stokes and anti-Stokes lines are witten by dashed lines and solid 
lines, respectively. \label{Stokes} }
\end{figure}
%%%%%%%%%%%%%%%%%%%%

We should consider the problem concerning the ``Stokes phenomenon" related 
to the zeros and poles of $Q^{2}$ \cite{Berry}, which are written 
in Fig.~\ref{Stokes} in the limit Im$(\omega )\to -\infty$. 
One of the important points are that the zeros of $Q^2$ approach 
the origin in the limit Im$(\omega)\to -\infty$. 
Near the origin, we can write as 
%%%%%%%%%%%%%%
\begin{eqnarray}
Q^{2}=g^{-2}\left[\omega^2 -\frac{4g^2}{r^2}\right]\ .\label{q}
\end{eqnarray}
%%%%%%%%%%%%%%%
Since $g\to -2M/r$ for $r\to 0$ where $M$ is the mass of Schwarzschild black hole, 
$Q^{2}$ has four zeros. When we start the outgoing solution at the point $a$ as 
%%%%%%%%%%%%%%%%
\beqa
\Psi_{a}=\Psi_{1}^{(r_{1})}\ , \label{pointa}
\eeqa
%%%%%%%%%%%%%%%
and proceeds along anti-Stokes lines and encircles the pole at the horizon clockwise, 
and turns back to $a$, we investigate what conditions are imposed to reproduce the 
original solution (\ref{pointa}). For this purpose, we should account for 
the Stokes phenomenon associated with the zeros $r_{1}$, $r_{2}$ and $r_{3}$. 
For example, if we proceeds the point $a$ to $b$ passing the Stokes line, 
we have the solution 
%%%%%%%%%%%%%%%%
\beqa
\Psi_{b}=e^{-iI}\Psi_{1}^{(r_{3})}-ie^{iI}\Psi_{2}^{(r_{3})}\ , \label{pointb}
\eeqa
%%%%%%%%%%%%%%%
where 
%%%%%%%%%%%%%%
\begin{eqnarray}
I=\int_{r_{3}}^{r_{1}}Qdr\ .
\end{eqnarray}
%%%%%%%%%%%%%%%
For details, see \cite{Andersson}. The final condition to be imposed is 
%%%%%%%%%%%%%%
\begin{eqnarray}
e^{2i\Gamma}=-1-2\cos 2I \ , \label{condition}
\end{eqnarray}
%%%%%%%%%%%%%%%
where 
%%%%%%%%%%%%%%
\begin{eqnarray}
\Gamma =\oint Qdr\ .
\end{eqnarray}
%%%%%%%%%%%%%%%
We should also perform the same analysis for the ingoing solution near the event 
horizon. The result is same as (\ref{condition}). 

Let us evaluate $\Gamma$ and $I$. $\Gamma$ is written as 
%%%%%%%%%%%%%%
\begin{eqnarray}
\Gamma =-2\pi i\lim_{r\to r_{\rm H}}\frac{r-r_{\rm H}}{g}\sqrt{\omega^2 +(g')^{2}/4}\ ,
\end{eqnarray}
%%%%%%%%%%%%%%%
since the contributions from $V(r)$ and $-gg''/2$ disappear at the event 
horizon. Since the term $(g|_{r=r_{\rm H}}')^{2}/4
=4(\pi T_{\rm H})^2$ has finite value (Remember, (\ref{temperature}).), we can also neglect it 
in the limit Im$(\omega)\to -\infty$. Then, we have 
%%%%%%%%%%%%%%
\begin{eqnarray}
\Gamma &=&-2\pi i\lim_{r\to r_{\rm H}}\frac{r-r_{\rm H}}{g}\omega\ ,\nonumber \\ 
       &=&-2\pi i\lim_{r\to r_{\rm H}}\frac{r-r_{\rm H}}{g'(r-r_{\rm H})}\omega 
=-i\frac{\omega}{2T_{\rm H}}\ . \label{gamma}
\end{eqnarray}
%%%%%%%%%%%%%%%
Notice that this result does not depend on species of black holes which becomes 
important later. 

To integrate $I$, we define 
%%%%%%%%%%%%%%
\begin{eqnarray}
y=\frac{\omega r^2}{4M}\ . 
\end{eqnarray}
%%%%%%%%%%%%%%%
From (\ref{q}), we can perform the integral $I$ as 
%%%%%%%%%%%%%%
\begin{eqnarray}
I=-\int_{-1}^{1}\sqrt{1-\frac{1}{y^{2}}}dr=\pi\ . \label{int}
\end{eqnarray}
%%%%%%%%%%%%%%%
By substituting (\ref{gamma}) and (\ref{int}) into (\ref{condition}), 
we have (\ref{Key}) as derived in previous papers. 

Next, we consider generalization of the above argument by exmplifying the case 
in dilatonic black hole. The crux of the point we now show is that $Q^{2}(r)$ for 
dilatonic black holes have two second order poles and four zeros in the limit 
Im$(\omega)\to -\infty$ which is qualitatively same as Schwarzschild black hole. 
Dilatonic black hole can be expressed using the 
coordinate \cite{GM-GHS}
%%%%%%%%%%%%%%%
\begin{eqnarray}
ds^{2}=-\lambda^{2}(\rho) dt^{2}+
\frac{1}{\lambda^{2} }d\rho^{2}+r^{2}(\rho)d\Omega^{2}\ , \label{forGM}
\end{eqnarray}
%%%%%%%%%%%%%%%
where 
%%%%%%%%%%%%%%%%
\begin{eqnarray}
\lambda^{2}&=& \left(1-\frac{\rho_{+}}{\rho}\right)
\left(1-\frac{\rho_{-}}{\rho}\right)^{(1-\alpha^{2})/(1+\alpha^{2})} \ ,
\label{lambdaGM}   \\
r&=& \rho\left(1-\frac{\rho_{-}}{\rho}\right)^{\alpha^{2}/(1+\alpha^{2})}\ .
\label{RGM}   
\end{eqnarray}
%%%%%%%%%%%%%%%
$\rho_{+}$, $\rho_{-}$ and $\alpha$ are the event horizon, the ``inner horizon", 
and the dilaton coupling, respectively. We can see from (\ref{RGM}) that 
the ``inner horizon" corresponds to the origin in the area radius. 

By comparing (\ref{forGM}) and (\ref{metric}), we obtain 
%%%%%%%%%%%%%%%
\begin{eqnarray}
g(r)&=&\left(1-\frac{\rho_{+}}{\rho}\right)\left(1-\frac{\rho_{-}}{\rho}
\right)^{1/(1+\alpha^{2})}\times \nonumber  \\
&&\left(1+\frac{\alpha^2}{1+\alpha^2}\frac{\rho_{-}}{\rho -\rho_{-}}\right)\ ,
\label{gr}
\\
e^{-\delta}&=&\left(1-\frac{\rho_{-}}{\rho}
\right)^{-\alpha^{2}/(1+\alpha^{2})}\times \nonumber \\
&&\left(1+\frac{\alpha^2}{1+\alpha^2}\frac{\rho_{-}}{\rho -\rho_{-}}\right)^{-1}\ .
\label{delta}
\end{eqnarray}
%%%%%%%%%%%%%%%
At first glance, it is not evident whether or not zeros of $Q^{2}$ approach the origin 
in the limit Im$(\omega)\to -\infty$. However, we can find from (\ref{gr}) and (\ref{delta}) 
that $e^{-\delta}$ and $g(r) $ do not show singular behavior for $r\neq 0$, $r_{\rm H}$ 
($\rho\neq\rho_{-}$, $\rho_{+}$) as it is expected from the fact that dilatonic black hole 
is a single-horizon black hole. Thus, zeros approaches the origin as in the Schwarzschild 
case. We evaluate $g(r)$ in the limit $r\to 0$, which is 
%%%%%%%%%%%%%%%
\begin{eqnarray}
g(r)\simeq \frac{\alpha^2}{1+\alpha^2}\frac{\rho_{-} -\rho_{+}}{
(\rho -\rho_{-})^{\frac{\alpha^2}{1+\alpha^2}} \rho^{\frac{1}{1+\alpha^2} } }\ .
\end{eqnarray}
%%%%%%%%%%%%%%%
If we substitute (\ref{RGM}) in this relation, we obtain 
%%%%%%%%%%%%%%
\begin{eqnarray}
g(r)\simeq \frac{\alpha^2}{1+\alpha^2}\frac{\rho_{-} -\rho_{+}}{r}\ . \label{GMgr}
\end{eqnarray}
%%%%%%%%%%%%%%%
Using this asymptotic relation to (\ref{r}), we have $Q^{2}(r)=R-1/(4r^2)$ again 
for $\Psi$ to behave near the origin appropriately. Then, we have the form 
(\ref{q}) near the origin and using the fact that dilatonic black 
hole has one horizon, we find that $Q^{2}(r)$ have four zeros and two second 
order poles as in Schwarzschild black hole.  

Therefore, the WKB condition to obtain the global solution is quite analogous to the 
case in Schwarzschild black hole and is written as (\ref{condition}). 
As we noted above, the expression (\ref{gamma}) is not also changed in 
dilatonic black hole. The nontrivial factor is $I$. However, since only difference of $g(r)$ 
in (\ref{GMgr}) from Schwarzschild case is its coefficient, if we define 
%%%%%%%%%%%%%%
\begin{eqnarray}
y=\frac{\omega r^2}{\frac{2\alpha^2}{1+\alpha^2}(\rho_{+} -\rho_{-})}\ , 
\end{eqnarray}
%%%%%%%%%%%%%%%
we can also perform the integral $I$ as (\ref{int}). Thus, we obtain 
(\ref{Key}) again which is the realization of the conjecture in \cite{Visser,Pad}. 

As for scalar and electromagnetic perturbations, we can perform them quite 
analogously. Using the asymptotic behavior 
%%%%%%%%%%%%%%
\begin{eqnarray}
R(r)\simeq \frac{(1-k)(1+k)}{r^2}-\frac{3}{4r^2}\ , \label{asym2}
\end{eqnarray}
%%%%%%%%%%%%%%%
in the limit $r\to 0$, we obtain 
%%%%%%%%%%%%%%
\begin{eqnarray}
e^{2i\Gamma}=-1-2\cos 2\pi k \ , \label{condition2}
\end{eqnarray}
%%%%%%%%%%%%%%%
where $k=0$, $1$ for scalar and electromagnetic perturbations, respectively. 
Thus, (\ref{Key}) also holds for scalar perturbations and the real part of 
electromagnetic perturbations disappears as for the case in 
Schwarzschild black hole. 

For even parity gravitational perturbations of dilatonic black hole, isospectrality 
between odd and even parity mode does not hold and the corresponding basic equation 
becomes complicated as shown in Ref.~\cite{Ferrari}. However, there remains a 
possibility that isospectrality is restored in the highly damped mode. 
This is under investigation. 

From the observation for the case in dilatonic black hole,  
the important things are: (i) the number of poles in $Q^2$ which is restricted to two 
in the single-horizon black holes. (ii) the number of zeros in $Q^2$ near the origin. 
(iii) asymptotically flatness that guarantees our boundary conditions. 
Therefore, if we turn back the case for higher dimensional Schwarzschild black hole 
in Ref.~\cite{Motl,Kuns,Birm,Lemos}, it is not difficult to extend the 
formula (\ref{Key}) for single-horizon black 
holes which behave near the origin as 
%%%%%%%%%%%%%%
\begin{eqnarray}
g(r)\simeq \frac{C}{r^{n}}\ , \label{generalgr}
\end{eqnarray}
%%%%%%%%%%%%%%%
where $C$ and $n$ are the constant and the natural number, respectively. 
Unfortunately, since black holes with non-Abelian fields, which have one horizon 
in general, show complicated 
behavior near the origin \cite{Gal'tsov,Maison,Higgs,dilaton,Tamaki}, 
we need further analysis to include these cases. 

%%%%%%%%%%%%%%%%%%%%%%%%%
{\it Conclusion and discussion. }
%%%%%%%%%%%%%%%%%%%%%%%%%
We investigated the highly damped quasinormal mode of single-horizon 
black holes and obtained the relation (\ref{Key}) for dilatonic 
black hole and considered the possibility of its generality. 
Our results are important since we supply the first example which shows 
(\ref{Key}) for black holes with matter fields. They suggest the generality 
of (\ref{Key}) in single-horizon black holes. Then, what we think about 
the confrontation in determining the Immirzi parameter $\gamma$ and the case 
in multi-horizon black holes ? It would be worth examining the 
present proposals \cite{Alekseev,Corichi2,Corichi3} since 
the results $j_{\rm min}$ and $\gamma$ in both cases (would) turn out 
to be general for single-horizon black holes, and are too close to ignore and 
suggest some relations. 

First, the possibility of modified area spectrum in Ref.~\cite{Alekseev} 
is not correct. Notice that the physical state does not change by 
adding or removing closed loops with $j=0$. The problem is that $j=0$ spin network 
has nonzero eigenvalue for the area operator. That is, we can obtain different 
eigenvalues for the area to the same physical state \cite{Corichi3}. 
Thus, we cannot accept this possibility. 

The mechanism that prohibits the transition $j=1/2$ by the fermion conservation 
is important \cite{Corichi2}. This implies $j_{\rm min}=1$ if we consider the 
{\it dynamical} process in the area change. However, we should recongnize that 
$j_{\rm min}$ in (\ref{SLQG}) means the statistically dominant element which 
does not necessarily coincide with the former. 
The drawback in Ref.~\cite{Corichi2} is that we can not 
prohibits the existence of $j=1/2$ edges puncturing the horizon 
as it was already pointed out. 
Therefore, it is important to investigate the mechanisim that suppress 
(or prohibit) $j=1/2$ punctures. For the supersymmetric case, this mechanism 
{\it would} exist as discussed in Ref.~\cite{Ling}. 

However, there is another possibility. 
In our opinion, the discussion in the quasinormal modes is like the old quantum 
theory and its description is within the general relativity. 
Thus, the above confrontation and the appearent discrepancy for multi-horizon 
black holes may be caused by this temporal description. If we can appropriately 
consider the problem corresponding to the quasinormal modes in 
the LQG, these may be solved. It is one of the directions we are seeking for. 

It is also important to consider other correspondence as done in BTZ black 
hole in Ref.~\cite{BTZ}. In this case, identification of the real part of 
the quasinormal frequencies with the fundamental quanta of black hole mass and 
angular momentum leads to the quantum behavior of the asymptotic symmetry algebra. 
At present, their relation to the loop quantum gravity is not clear. 
It is also the important direction we should seek for. 

Of course, there are problems we should solve before going to the consideration above. 
We need to prove the case for single-horizon black holes in possibly 
general form. We must also include the case for the even parity mode.  
They are the work we are now considering. 

%%%%%%%%%%%%%%%%%%%%%%%%%
{\it Acknowledgement.}
%%%%%%%%%%%%%%%%%%%%%%%%%
Special thanks to Hajime Sotani, Jiro Soda, Shijun Yoshida, and Yoshiyuki Morisawa 
for useful discussion. 
This work was supported in part by Grant-in-Aid for Scientific Research Fund of the
Ministry of Education, Science, Culture and Technology of Japan, 2003, No.\ 154568
(T.T.).  This work was also supported in part by a
Grant-in-Aid for the 21st Century COE ``Center for Diversity and Universality in 
Physics".

%%%%%%%%%%%%%%%%%%%%%%%%%%%%%%%%%%%%%%%%%%%%%%%%%%

%%%%%%%%%%%%%%%%%%%%%%%%%%%%%%%%%%%%%%%%%%%%%%%

\end{document}